\documentstyle[11pt,newpasp,epsf]{article}

\begin{document}

\title{A Magnetic Betelgeuse? Numerical Simulations of Non-linear Dynamo Action}

\author{S.B.F. Dorch} 
\affil{The Niels Bohr Institute, Juliane Maries Vej 30, DK-2100 Copenhagen} 

\setcounter{footnote}{0}

\begin{abstract}
Betelgeuse is an example of a cool super-giant displaying brightness 
fluctuations and irregular surface structures. 
Simulations by Freytag, Steffen, \& (2002) of the convective
envelope of the star have shown that the fluctuations 
in the star's luminosity may be caused by giant cell 
convection. A related question regarding the nature of Betelgeuse
and supergiants in general is whether these stars may be 
magnetically active. If so, that
may in turn also contribute to their variability.
By performing detailed numerical simulations,
I find that both linear kinematic and non-linear dynamo action are possible
and that the non-linear magnetic field saturates at a value 
somewhat below equipartition: in the linear regime there are two modes 
of dynamo action.
\end{abstract}


\index{*Betelgeuse|alpha Ori}

\section{Introduction}

The cool super-giant star Betelgeuse is one of the 
the stars with the largest apparent diameters on the 
sky---corresponding 
to a radius somewhere in the range 600--800 ${\rm R}_{\odot}$.
Freytag, Steffen, \& Dorch 2002 performed detailed numerical 
three-dimensional radiation-hydrodynamic simulations of the 
outer convective envelope and atmosphere of the
star under realistic physical assumptions. They tried to determine if 
its observed brightness variations may be understood as convective 
motions within the star's atmosphere: the resulting models are 
largely successful in explaining the observations as a consequence 
of giant-cell convection on the stellar surface, very dissimilar 
to solar convection. These detailed simulations bring forth the 
possibility of solving another question regarding the nature of 
Betelgeuse and super-giants in general; namely whether such 
stars may harbor magnetic activity that in turn may also 
contribute to their variability and other phenomena derived from
the presence of a magnetic field (such as dust formation
and mass-loss). A possible astrophysical dynamo 
in Betelgeuse would most likely be very different from those 
thought to operate in solar type stars, both due to its slow 
rotation, and to the fact that only a few convection cells are 
present at its surface at any one time. 

\section{Model}

We solve the full three-dimensional magneto-hydrodynamical (MHD) equations 
for a fully convective star---a so 
called ``star-in-a-box'' simulation---employing the ``Pencil Code'' by 
Brandenburg \& Dobler (2003). This code has a specially designed ``convective 
star'' module, that allows the solution of the non-linear MHD equations
by the numerical pencil scheme in a star with a fixed radius R and mass M. 
Variables are measured in terms of these latter two parameters so that e.g.\ the
unit of the star's luminosity L becomes 
$\frac{\rm M}{\rm R} ({\rm GM}/{\rm R})^{\frac{3}{2}}$. 
In the present case I set ${\rm R} = 640~ {\rm R}_{\odot}$ 
and ${\rm M} = 5~ {\rm M}_{\odot}$ yielding a luminosity of 
${\rm L} = 46000~ {\rm L}_{\odot}$, consistent with current estimates 
of Betelgeuse' actual size, mass and luminosity. 
We use a numerical resolution
of $128^3$ uniformly distributed grid points, a fixed gravitational potential,
an inner small heating core, and an outer thin isothermally cooling surface 
at $r = {\rm R}$, with a cooling time scale set to $\tau_{\rm cool} = 1$ year
(corresponding to the model of Freytag et al. 2002).

Betelgeuse is only slowly rotating and a rotational 
frequency was chosen corresponding to a surface rotational velocity of 
5 km/s, yielding a small inverse Rossby number
meaning that the flows are not very helical. 

\subsection{Diffusion and magnetic Reynolds number}

Dynamo action by flows are often studied in the limit of 
increasingly large magnetic Reynolds numbers Re$_{\rm m} = \ell 
{\rm U}/\eta$, where $\ell$ and U are characteristic length and 
velocity scales. Most astrophysical systems are highly conducting 
yielding small magnetic diffusivities $\eta$, and 
their dimensions are huge resulting in huge values of Re$_{\rm m}$. 
Betelgeuse is not an exception; 
most parts of the star is better conducting than the solar surface 
layers, which has a magnetic diffusivity of the order of $\eta 
\approx 10^4$ m$^2$/s (see Dorch \& Freytag 2002). 

There is some uncertainty as to defining the most 
important length scale of the system, but taking $\ell$ to be 10\% 
of the radial distance from the center (a typical scale of the 
giant cells), and U $= {\rm u}_{\rm RMS}$ along the radial direction 
yields Re$_{\rm m}=10^{10}$--$10^{12}$ in the interior part of the 
star where R $\le 700$ R$_{\odot}$. 

In the present case we cannot use that large values of Re$_{\rm m}$
(partly due to the fact that we are employing a uniform fixed diffusivity),
but rely on the results from generic dynamo simulations indicating 
that results converge already at Reynolds numbers of a few hundred
(e.g.\ Archontis, Dorch, \& Nordlund 2003a, 2003b).
Furthermore, Dorch \& Freytag (2002) obtained kinematic dynamo action
in their model of a magnetic Betelgeuse at Re$_{\rm m} \sim 500$.
Here we put Re$_{\rm m} \sim 300$ (based on the largest scales).
Additionally the magnetic Prantl number is Pr$_{\rm m} \sim 80$.

\begin{figure}[!htb]
\plotone{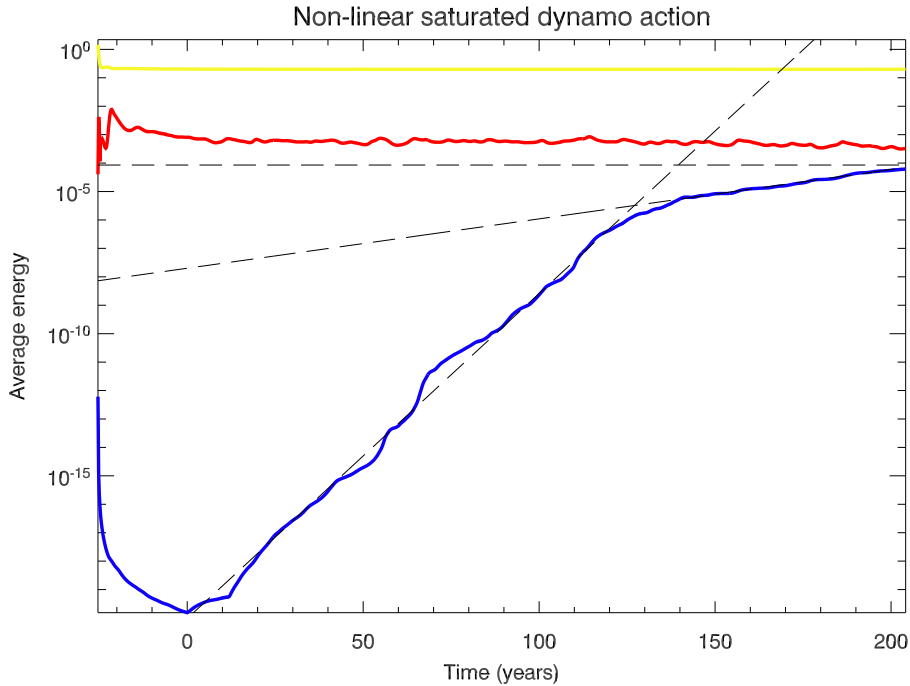} 
\caption{Linear regime: energy as a function of Betelgeusian time in 
years. Upper (almost horizontal) line is total thermal energy E$_{\rm th}$
(yellow in color print), middle full horizontal curve with wiggles is total kinetic 
energy E$_{\rm kin}$ (red in color print), and lower full curve is E$_{\rm m}$
(blue in color print). The dashed thin lines indicate growth corresponding to 
growth times of 3.8, 25 and $\infty$ years.} \label{fig-1}
\end{figure}

\begin{figure}[!htb]
\plotone{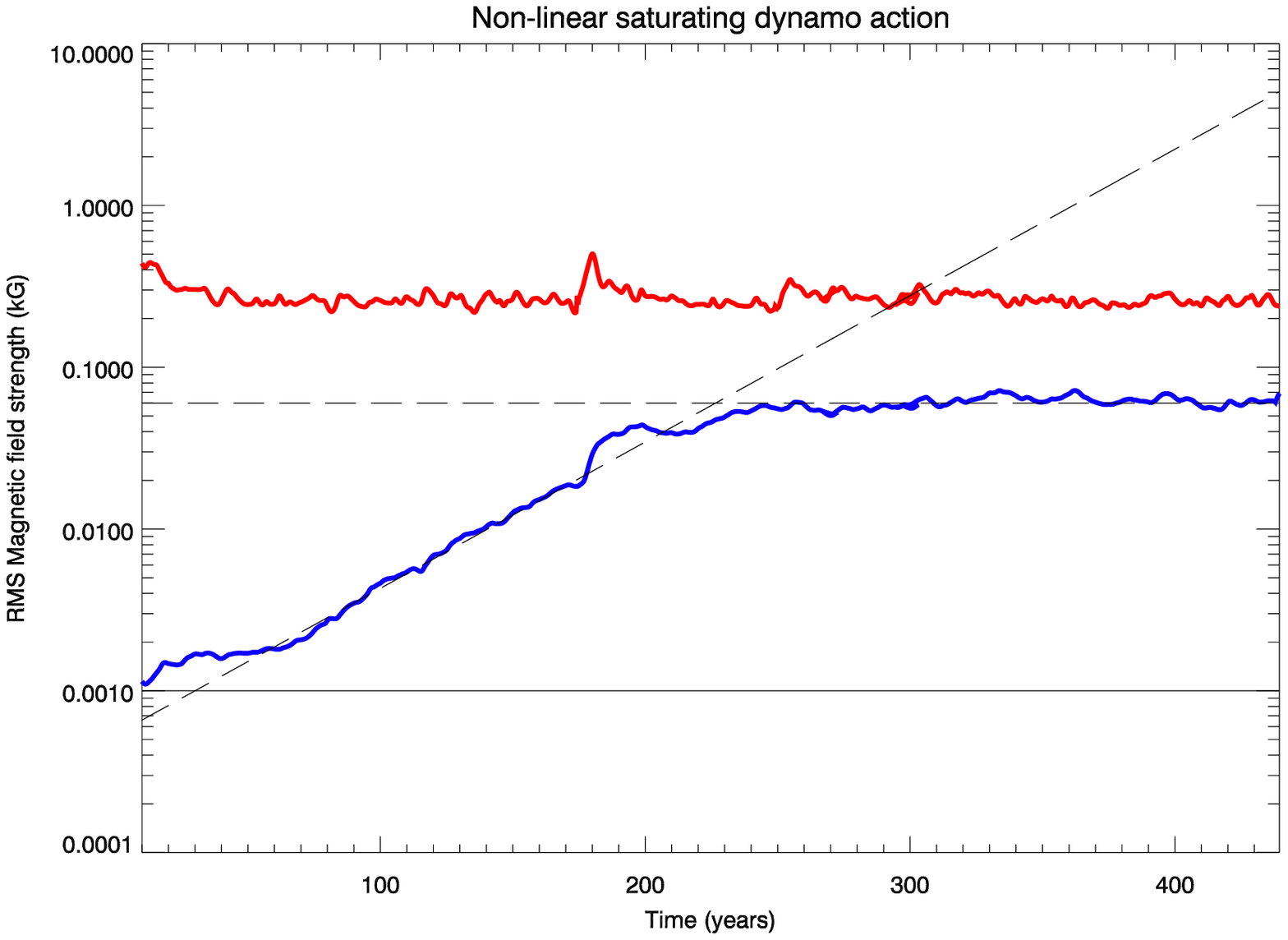} 
\caption{Transition to the non-linear regime:
RMS magnetic field B in kilo-Gauss (kG) as a
function of Betelgeusian time in years. Upper full curve is the
equipartition field strength B$_{\rm eq}$ corresponding to the fluid motions 
(red in color print)
and the lower full curve is the actual RMS field strength B (blue in color print).
The dashed thin lines correspond to growth times of 25 and $\infty$ years,
while the full thin line is the field strength at $t = 10~$ year, corresponding
approximately to the start of the second linear mode.} \label{fig-2} 
\end{figure}

\section{Results}

There is some disagreement as to what one should require 
for a system to be an astrophysical dynamo. Several ingredients seems 
necessary: flows must stretch, twist and fold the magnetic field lines;
reconnection must take place to render the processes irreversible; 
weak magnetic field must be circulated to the locations where 
flow can do work upon it; and finally,
the total volume magnetic energy E$_{\rm m}$ must increase (kinematic regime)
or remain at constant amplitude on a long time scale (non-linear regime).
These points are based largely on experience from idealized kinematic and 
non-linear dynamo models; e.g.\ Archontis et al. (2003a, 2003b). 
Here we shall deal mainly with the question of the exponential growth 
and saturation of of E$_{\rm mag}$. 

In an earlier kinematic study of Betelgeuse using a completely 
different numerical approach (Freytag et al.\ 2002 and
Freytag \& Dorch 2002), dynamo action was 
obtained when the specified minimum value of 
Re$_{\rm m}$ was larger than approximately 500 (at lower values of 
Re$_{\rm m}$ the total magnetic energy decayed). In the present 
case with Re$_{\rm m} \sim 300$, we find an initial clear exponential 
growth over several turn-over times, and many orders of magnitude in energy.
Figure \ref{fig-1} shows the evolution of E$_{\rm mag}$ as a function of time,
for the first 200+ years (in Betelgeusian time): initially there is a ``short''
transient of 25 years, where the field exponentially decays because the fluid
motions has not yet attained their final amplitude. 
Once the giant cell convection has properly
begun, however, the magnetic field is amplified and we enter a linear regime
(of exponential growth). There are two modes of amplification
in the linear regime; the initial mode with a growth rate of about 3.8 years,
which in the end gives way to a mode with a smaller growth rate corresponding
to 25 years. This is a slightly strange situation, since normally
modes with smaller growth rates are overtaken by modes with larger growth rates
(cf.\ Dorch 2000); the explanation is probably that while both modes
are growing modes, only the one with the largest growth rate is a purely
kinematic mode---the second mode is not kinematic, but the presence of the
magnetic field is felt by the fluid (through the Lorentz force becoming
important), which quenches the growth slightly.

No exponential growth can go on forever and eventually the magnetic energy
amplification must saturate. Figure \ref{fig-2} shows E$_{\rm mag}$ as a function 
of time for 440 Betelgeusian years: the second linear mode as well as
the mode in the non-linear regime are shown. The magnetic field saturates at an
RMS values of about 60 Gauss, corresponding to a total magnetic energy E$_{\rm mag}$
below equipartition with the kinetic energy E$_{\rm kin}$ by a factor of $\sim 0.25$.

We can study the geometry of the magnetic field that the 
dynamo generates\footnote{The CD-ROM contains a volume rendering of 
isosurfaces of magnetic field structures with a high field strength relative
to the maximum: the figure {\tt 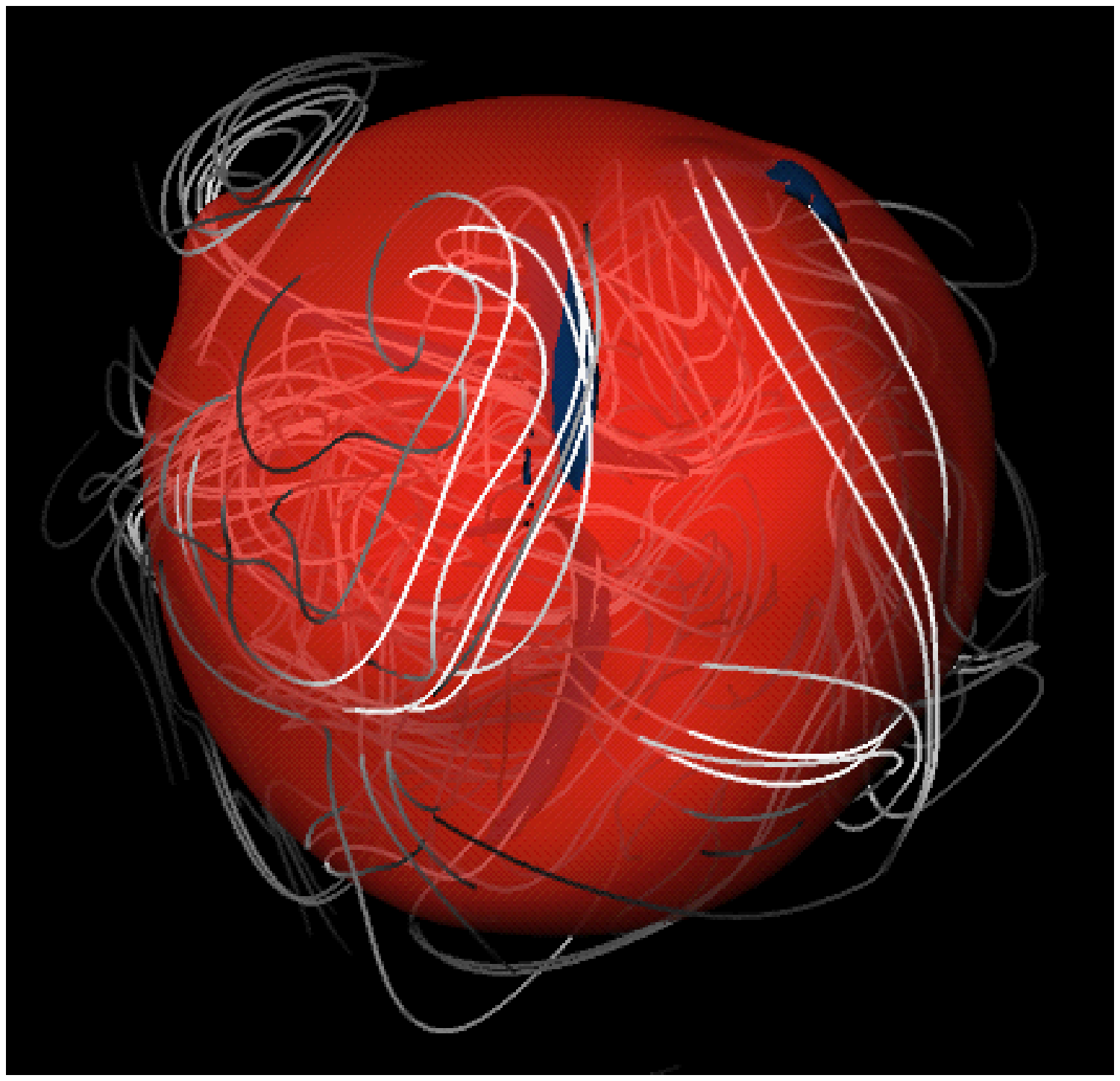}.}: 
the field becomes concentrated into elongated structures 
much thinner than
the scale of the giant convection cells, but perhaps due to the very dynamic
nature of the convective flows, no ``intergranular network'' is formed 
in the solar sense. The field is highly intermittent,
i.e.\ only a small fraction of the volume carries the
strongest structures. 
The magnetic structures are well resolved, with maximum
power on the largest scales corresponding to wavenumbers of a few.
Additionally we observe a slight trend in the topology of the
field (see also Freytag \& Dorch 2002); 
fields near the surface of the star are predominantly
horizontally aligned, while those in deeper layers are more or less radial.

\section{Concluding remarks}

Based on the results presented here, we may not say conclusively if 
Betelgeuse has a magnetic field;
the results are tentative and should be used with caution.
But we may say that it seems that it might
have a presently unobserved magnetic field. 
The main highlights are the following:
\begin{itemize}
 \item In the linear regime two modes are present: an initial kinematic mode
  with a high growth rate is overtaken by a linear mode with a lower growth
  rate and thus a longer growth time.
 \item The growth time of the last occurring mode in the linear regime
  is about 25 years, i.e.\ the same value as was found in previous purely 
  kinematic dynamo models (Freytag et al.\ 2002).
 \item In the non-linear regime the field strength saturates at an RMS 
  value of about 60 Gauss, corresponding to sub-equipartition at 
  ${\rm E}_{\rm mag} \approx 0.25~ {\rm E}_{\rm kin}$.
 \item Magnetic structures in the non-linear regime are large by solar
  standards, but smaller than the giant convection cells, with a typical
  scale of about 15\% of the radius.
\end{itemize}

\acknowledgments

SBFD thanks the LOC of IAU Symposium No.\ 219 for accepting this
poster presentation for publication in the proceedings. SBFD
is supported by a Steno grant from the Danish Natural Science
Research Council. The computer simulations were possible thanks
to the establishment of the Danish Center for Scientific
Computing in particular the Horseshoe cluster at Odense
University.

\end{document}